\documentclass[copyright,creativecommons]{eptcs}
\providecommand{\event}{QAPL 2015} 
\usepackage{breakurl}             

\usepackage{amsfonts}
\usepackage{amsmath}
\usepackage{stmaryrd}
\usepackage{mathtools}
\usepackage{proof}
\usepackage[T1]{fontenc}
\title{\caspa{}: Collective Adaptive Resource-sharing Markovian Agents}
\usepackage{xcolor}
\newcommand{\caspa}{\textsc{Carma}}
\newcommand{\csnets}{\textsc{Col}}

\newcommand{\cssystems}{\textsc{Sys}}
\newcommand{\cscomps}{\textsc{Comp}}
\newcommand{\csprocesses}{\textsc{Proc}}
\newcommand{\Labs}{\textsc{Lab}}

\newcommand{\csnil}{\mathbf{nil}}
\newcommand{\cskill}{\mathbf{kill}}
\newcommand{\csguard}[2]{[#1]#2}

\newcommand{\csppar}{~|~}
\newcommand{\cscpar}{\parallel}
\newcommand{\cszero}{\mathbf{0}}

\newcommand{\broadcastsymbol}{\star}

\newcommand{\csgout}[5]{#2^{#1}[ #3 ]\langle #4\rangle #5 }
\newcommand{\csbout}[4]{\csgout{\broadcastsymbol}{#1}{#2}{#3}{#4}}
\newcommand{\csout}[4]{\csgout{}{#1}{#2}{#3}{#4}}
\newcommand{\csgin}[5]{#2^{#1}[ #3 ]( #4 ) #5 }
\newcommand{\csbin}[4]{\csgin{\broadcastsymbol}{#1}{#2}{#3}{#4}}
\newcommand{\csin}[4]{\csgin{}{#1}{#2}{#3}{#4}}

\newcommand{\labgout}[5]{#2^{#1}[#3]\langle #4\rangle,#5}
\newcommand{\labgin}[5]{#2^{#1}[#3]( #4 ),#5}
\newcommand{\labbout}[4]{\labgout{\broadcastsymbol}{#1}{#2}{#3}{#4}}
\newcommand{\labbin}[4]{\labgin{\broadcastsymbol}{#1}{#2}{#3}{#4}}
\newcommand{\labout}[4]{\labgout{}{#1}{#2}{#3}{#4}}
\newcommand{\labin}[4]{\labgin{}{#1}{#2}{#3}{#4}}
\newcommand{\labsync}[4]{\tau[\labgout{}{#1}{#2}{#3}{#4}]}
\newcommand{\refusal}[4]{\mathcal{R}[\labgin{\broadcastsymbol}{#1}{#2}{#3}{#4}]}

\newcommand{\cstrue}{\top}
\newcommand{\csfalse}{\bot}

\newcommand{\csupd}{\sigma}
\newcommand{\csstore}{\gamma}
\newcommand{\col}{N}
\newcommand{\comp}{C}
\newcommand{\cscomp}[2]{(#1,#2)}
\newcommand{\sys}{S}
\newcommand{\cssys}[2]{#1~\mathbf{in}~#2}
\newcommand{\csenv}{\mathcal{E}}

\newcommand{\cssel}{\mu}

\newcommand{\guard}{\pi}

\newcommand{\seq}[1]{\overrightarrow{#1}}

\newcommand{\defi}{\stackrel{\mathit{\triangle}}{=}}

\newcommand{\ActT}{\mbox{\textsc{ActType}}}
\newcommand{\Act}{\mbox{\textsc{Act}}}
\newcommand{\Attr}{\mbox{\textsc{{Attr}}}}
\newcommand{\Var}{\mbox{\textsc{Var}}}
\newcommand{\Val}{\mbox{\textsc{{Val}}}}

\newcommand{\comparrow}[2]{\xrightharpoondown{#1}_{#2}}
\newcommand{\colarrow}[2]{\xrightarrow{#1}_{#2}}
\newcommand{\sysarrow}[2]{\xmapsto{#1}}

\usepackage{todonotes}

\newcommand{\rulename}[1]{\mbox{\textbf{#1}}}

\newcommand{\modif}[1]{{\color{black}{#1}}}

\author{
Luca Bortolussi
\institute{Saarland University\\University of Trieste\\ISTI - CNR\\}
\and
Rocco De Nicola
\institute{IMT Lucca}
\and
Vashti Galpin
\institute{University of Edinburgh}
\and 
Stephen Gilmore
\institute{University of Edinburgh}
\and
Jane Hillston
\institute{University of Edinburgh}
\and
Diego Latella
\institute{ISTI - CNR}
\and
Michele Loreti
\institute{Universit\`a di Firenze\\IMT Lucca}
\and
Mieke Massink
\institute{ISTI - CNR}
}
\def\titlerunning{\caspa{}: Collective Adaptive Resource-sharing Markovian Agents}
\def\authorrunning{L.~Bortolussi \emph{et al.}}

\begin{document}
\maketitle

\begin{abstract}
In this paper we present \caspa{}, a language recently defined to support specification and analysis of collective adaptive 
systems.
\caspa{} is a stochastic process algebra equipped with linguistic constructs 
specifically developed for modelling and programming systems that can operate in 
open-ended and unpredictable environments. 
This class of systems is typically composed of a huge number of interacting 
agents that dynamically adjust and combine their behaviour to achieve specific 
goals. 
A \caspa{} model, termed a \emph{collective}, consists of a set of components, each of which 
exhibits a set of attributes. To model dynamic aggregations, which are sometimes referred to
as \emph{ensembles}, \caspa{} provides communication primitives that are based on 
predicates over the exhibited attributes. These predicates are used to select the participants in a communication. 
Two communication mechanisms are provided in the \caspa{} language: multicast-based and unicast-based.  
In this paper, we first 
introduce the basic principles of \caspa{} and then we show how our language can be 
used to support specification with a simple but illustrative example of a socio-technical collective adaptive system.
\end{abstract}

\section{Introduction}
\label{sec:introduction}

\emph{Collective adaptive systems} (CAS) typically consist of very large numbers of components which exhibit autonomic behaviour depending on their properties, objectives and actions. Decision-making in such systems is complicated and interaction between their components may introduce new and sometimes unexpected behaviours. CAS are open, in the sense that components may enter or leave the collective at anytime. Components can be highly heterogeneous (machines, humans, networks, etc.) each operating at different temporal and spatial scales, and having different (potentially conflicting) objectives. We are still far from being able to design and engineer real collective adaptive systems, or even specify the principles by which they should operate. 

CAS thus provide a significant research challenge in terms of both representation and reasoning about their behaviour. The pervasive yet transparent nature of  the applications developed in this paradigm makes it of paramount importance that their behaviour can be thoroughly assessed during their design, prior to deployment, and throughout their lifetime.  Indeed their adaptive nature makes modelling essential and models play a central role in driving their adaptation.  Moreover, the analysis should encompass both functional and non-functional aspects of behaviour.  Thus it is vital that we have available robust modelling techniques which are able to describe such systems and to reason about their behaviour in both qualitative and quantitative terms.  
To move towards this goal,  we consider it important to develop a theoretical foundation for collective adaptive systems that would help in understanding their distinctive features.
In this paper we present \caspa{}, a language designed within the
QUANTICOL project\footnote{\url{http://www.quanticol.eu}} specifically for the specification and analysis of CAS, with the particular objective of supporting quantitive evaluation and verification.  


\caspa{} builds on a long tradition of stochastic process algebras such as PEPA~\cite{hillston1995compositional}, MTIPP~\cite{mtipp}, EMPA~\cite{empa}, Stochastic $\pi$-Calculus~\cite{priami1995stochastic}, Bio-PEPA~\cite{ciocchetta2009bio}, \modif{MODEST~\cite{bohnenkampDHK06}} and others~\cite{HermannsHK02,BP10}.  It combines the lessons which have been learned from these languages with those learned from developing languages to model CAS, such as SCEL~\cite{NicolaLPT14} and PALOMA~\cite{palomaqest14}, which feature attribute-based communication and explicit representation of locations.  

SCEL~\cite{NicolaLPT14}  (Software Component Ensemble Language), is a kernel language that has been designed to support the programming of autonomic computing systems. This language relies on the notions of \emph{autonomic components} representing the collective members, and \emph{autonomic-component ensembles} representing collectives. Each component is equipped with an interface, consisting of a collection of attributes, describing different features of components. Attributes are used by components to dynamically organise themselves into ensembles and as a means to select partners for interaction. The stochastic variant of SCEL, called StocS \cite{LLMS14}, was a first step towards the investigation of the impact of different stochastic semantics for autonomic processes, that relies on stochastic output semantics, probabilistic input semantics and on a probabilistic notion of knowledge.  Moreover, SCEL has inspired the development  of the core calculus AbC~\cite{SAC15} that focuses on a minimal set of primitives that defines attribute-based communication, and investigates their impact. Communication among components takes place in a broadcast fashion, with the characteristic that only components satisfying predicates over specific attributes receive the sent messages, provided that they are willing to do so.

PALOMA~\cite{palomaqest14} is a process algebra that takes as starting point a model based on located Markovian agents each of which is parameterised by a location, which can be regarded as an attribute of the agent. The ability of agents to communicate depends on their location, through a perception function. This can be regarded as an example of a more general class of attribute-based communication mechanisms. The communication is based on a multicast,  as only agents who enable the appropriate reception  action have the ability to receive the message. 
The scope of  communication is thus adjusted according to the perception function. 

A distinctive contribution of the new language is the rich set of communication primitives that are offered.  \caspa{} supports both unicast and broadcast communication, and locally synchronous, but globally asynchronous communication.  This richness is important to enable the spatially distributed nature of CAS, where agents may have only local awareness of the system, yet the design objectives and adaptation goals are often expressed in terms of global behaviour.  Representing these rich patterns of communication in classical process algebras or traditional stochastic process algebras would be difficult, and would require the introduction of additional model components to represent buffers, queues and other communication structures.  
Another feature of \caspa{} is the explicit representation of the environment in which processes interact, allowing rapid testing of a system under different
open world scenarios. The environment in \caspa{} models can evolve at runtime, due to the feedback from the system, and it further modulates the interaction
between components, by shaping rates and interaction probabilities.
Furthermore the large scale nature of CAS systems
makes it essential to support scalable analysis techniques, thus \caspa{} has been designed anticipating both a discrete and a continuous semantics in the style of \cite{tribastone2012scalable}.

The focus of this paper is the presentation of the language and its discrete semantics, which are presented in the
\textsc{FuTS} style~\cite{DLLM13}.  The structure of the paper is as follows.  Section~2 presents the syntax of the language and explains the organisation of a model in terms of a collective of agents that are considered in the context of an environment.  In Section~3 we give a detailed account of the semantics, particularly explaining the role of the environment.   The use of \caspa{}  is illustrated in Section~4 where we describe a model of a simple bike sharing system.  Some conclusions are drawn in Section~5.

\section{\caspa{} syntax}
\label{sec:caspasyntax}

A \caspa{} system consists of a \emph{collective} ($\col$) operating in an \emph{environment} ($\csenv$). The collective consists of 
a set of components.  It models the behavioural part of a system and is used to describe a set of interacting \emph{agents} that cooperate
to achieve a set of given tasks. The environment models all those aspects which are intrinsic to the context where
the agents under consideration are operating. The environment also mediates agent interactions.

We let \cssystems{} be the set of \caspa{} \emph{systems} \sys{} defined by the following syntax:
\[
\sys ~ ::= ~ \cssys{\col}{\csenv}
\]
where $\col$ is a collective and $\csenv$ is an environment. The latter provides the global state of the system and governs the interactions in the collective. 

We let \csnets{} be the set of collectives $\col$ which are generated by the following grammar:
\[
\begin{array}{rclr}
\col & ::= & \comp~~\big|~~\col\cscpar \col
\end{array}
\]

A collective $\col$ is either a  \emph{component} $\comp$ or the parallel composition of two collectives  ($\col\cscpar \col$).

A component $\comp$ can be either the \emph{inactive component}, which is denoted by $\cszero$, or a term of the form $\cscomp{P}{\csstore}$, where $P$ is a  \emph{process} and $\csstore$ is a \emph{store}.
A term $\cscomp{P}{\csstore}$ models an \emph{agent} operating in the system under consideration: the process $P$ represents the agent's behaviour whereas the store $\csstore$ models its \emph{knowledge}.
A store is a function which maps \emph{attribute names} to \emph{basic values}. 
We let:
\begin{itemize}
\item $\Attr$ be the set of \emph{attribute names} $a$, $a'$, $a_1$,\ldots , $b$, $b'$, $b_1$,\ldots;
\item $\Val$ be the set of \emph{basic values} $v$, $v'$, $v_1$,\ldots;
\item $\Gamma$ be the set of \emph{stores} $\gamma,\gamma_1,\gamma',\ldots$ i.e.\ functions from 
$\Attr$ to $\Val$. 
\end{itemize}

We let $\cscomps$ be the set of components $\comp$ generated by the following grammar:
\[
\comp ::= \cszero~~\big|~~ \cscomp{P}{\csstore} 
\]

We let $\csprocesses$ be the set of processes $P$, $Q$,\ldots{} defined by the following grammar:

\begin{displaymath}
\begin{array}{c|c}
\begin{array}{rclr}
P,Q & ::=  & \csnil\\[2pt]
    & | & \cskill\\[2pt]
    & | & act.P\\[2pt]
    & | & P+Q\\[2pt]
    & | & P\csppar Q\\[2pt]
    & | & \csguard{\guard}{P}\\[2pt]
    & | & A & (A\defi P)\\
\end{array}
&    
\begin{array}{rclr}
act & ::= &  \csbout{\alpha}{\guard}{\seq{e}}{\csupd} & \mbox{} \\[4pt]
& | & \csout{\alpha}{\guard}{\seq{e}}{\csupd}\\[4pt]
& | & \csbin{\alpha}{\guard}{\seq{x}}{\csupd}\\[4pt]
& | & \csin{\alpha}{\guard}{\seq{x}}{\csupd}
\\[10pt]
e & ::= & a~|~\mathsf{this}.a~|~x~|~v~|~\cdots
\\[10pt]
\guard & ::= & \multicolumn{2}{l}{\cstrue~|~\csfalse~|~e_1\bowtie e_2~|~\neg\guard~|~\guard\wedge\guard~|~\cdots}
\end{array}
\end{array}
\end{displaymath}

In \caspa{} processes can perform four types of actions: \emph{broadcast output} ($\csbout{\alpha}{\guard}{\seq{e}}{\csupd}$),
\emph{broadcast input} ($\csbin{\alpha}{\guard}{\seq{x}}{\csupd}$), \emph{output} ($\csout{\alpha}{\guard}{\seq{e}}{\csupd}$),
and \emph{input} ($\csin{\alpha}{\guard}{\seq{x}}{\csupd}$).
Where:
\begin{itemize}
\item $\alpha$ is an \emph{action type} in the set of action type $\ActT$;
\item $\guard$ is an \emph{predicate};
\item $x$ is a \emph{variable} in the set of variables $\Var$;
\item $\seq{\cdot}$ indicates a sequence of elements;
\item $\csupd$ is an \emph{update}, i.e.\ a function from $\Gamma$ to $Dist(\Gamma)$
in the set of \emph{updates} $\Sigma$; where $Dist(\Gamma)$ is the set of probability distributions over $\Gamma$.
\end{itemize}

The admissible communication partners of each of these actions are identified by the predicate $\guard$. This is a predicate on \emph{attribute names}.
Note that, in a component  $\cscomp{P}{\csstore}$ the store $\csstore$ regulates the behaviour of $P$. Primarily, $\csstore{}$ is 
used to evaluate the predicate associated with an action in order to filter the possible synchronisations involving process $P$.
In addition, $\csstore$ is also used as one of the parameters for computing the actual rate of actions performed by $P$. 
The process $P$ can change $\gamma$ immediately after the execution of an action. 
This change is brought about by the \emph{update} $\sigma$. The update is a function that when given a store $\gamma$ returns a probability distribution over $\Gamma$ 
which expresses the possible evolutions of the store after the action execution.

The \emph{broadcast output} $\csbout{\alpha}{\guard}{\seq{e}}{\csupd}$ models the execution of an action $\alpha$ that spreads the values
resulting from the evaluation of expressions $\seq{e}$ in the local store $\csstore$. This message can be potentially received by any process
located at components whose store satisfies predicate $\guard$.
This predicate may contain references to attribute names that have to be evaluated under the local store. These references are prefixed by
the special name $\textsf{this}$.
For instance, if $\mathsf{loc}$ is the attribute used to store the position of a component, action
\[
\csbout{\alpha}{\mathsf{distance}(\mathsf{this}.\mathsf{loc},\mathsf{loc})\leq L}{\seq{v}}{\csupd}
\]
potentially involves all the components located at a distance that is less than or equal to a given threshold $L$.
The \emph{broadcast output} is non-blocking. The action is executed even if no process is able to receive the values which are sent.
Immediately after the execution of an action, the update $\csupd$ is used to compute the (possible) \emph{effects} of 
the performed action on the store of the hosting component where the output is performed.

To receive a broadcast message, a process executes a \emph{broadcast input} of the form $\csbin{\alpha}{\guard}{\seq{x}}{\csupd}$. 
This action is used to receive a tuple of values $\seq{v}$ sent with an action $\alpha$ from a component whose store satisfies the predicate 
$\guard[\seq{v}/\seq{x}]$. The transmitted values can be part of the predicate $\guard$.
For instance, $\csbin{\alpha}{x> 5}{x}{\csupd}$ can be used to receive a value that is greater than $5$.

The other two kinds of action, namely \emph{output} and \emph{input}, are similar. However, differently from broadcasts described
above, these actions realise a \emph{point-to-point} interaction. The \emph{output} operation is blocking, in contrast with the non-blocking broadcast output.

Choice and parallel composition are the usual definitions for process algebras.  Processes can be guarded so that $\csguard{\guard}{P}$ behaves
as the process $P$ if the predicate $\guard$ is satisfied.
Finally, process $\cskill$ is used to \emph{destroy} a component. We assume that this term always occurs under the scope of an action prefix.

%
%
%
%


\caspa{} collectives operate in an environment $\csenv{}$. This environment is used to model the intrinsic rules that govern,
for instance, the physical context where our system is situated. 

An environment consists of two elements: a \emph{global store} $\gamma_{g}$, that models the overall state of the system, and
an \emph{evolution rule} $\rho$. The latter is a function which, depending on the global store and the current state of the collective, i.e.\ the 
configurations of each component in the collective, returns a tuple of functions $\varepsilon=\langle \mu_{p}, \mu_{r}, \mu_{u}\rangle$ 
known as the \emph{evaluation context} where $\Act=\ActT\cup \{ \alpha^{\broadcastsymbol}| \alpha\in \ActT\}$ and:
\begin{itemize}
\item $\cssel_{p}: \Gamma\times\Act\rightarrow [0,1]$, expresses the probability to receive a message;
\item $\cssel_{r}: \Gamma\times\Gamma\times\Act\rightarrow \mathbb{R}_{\geq 0}$, computes the execution rate of an action;
\item $\cssel_{u}: \Gamma\times\Act\rightarrow \Sigma\times\csnets$, determines the updates on the environment (global store and collective) 
induced by the action execution. 
\end{itemize}

These functions regulate system behaviour. 
Function $\cssel_{p}$, which takes as parameters the local stores of the two interacting components, i.e.\ the sender and the receiver, and the action used to interact, returns the probability to receive a message. Function $\cssel_{r}$ computes the rate of an unicast/broadcast output. This function takes as parameter the local store of the component performing the action and the action on which interaction is based.  Note that the environment can disable the execution of a given action. This happens
when the function $\cssel_{r}$ (resp.\ $\cssel_{p}$) returns the value $0$.
Finally, the function $\cssel_{u}$ is used to update the global store and to install a new collective in the system. The function $\cssel_{u}$ takes as parameters the
store of the component performing the action together with the action type and returns a pair $(\csupd,\col)$.  Within this pair, $\csupd$ identifies the update on the global store whereas $\col$ is a new collective installed in the system. This function is particularly useful for modelling the arrival of new agents into a system. 
All of these functions are determined by an \emph{evolution rule} $\rho$ depending on the global store and the actual state of the components in the system.
For instance, the probability to receive a given message
may depend on the \emph{concentration} of components in a given state. Similarly, the actual rate of an action may be a function of the number of components
whose store satisfies a given property.

%
%
%

\section{\textsc{Carma} operational semantics}
\label{sec:caspasemantics}

In this section we define the operational semantics of \caspa{} specifications. This operational semantics is defined in three stages.
First, we introduce the transition relation $\comparrow{\ \ \ \cdot\ \ }{\cdot}$ that describes the behaviour of a single component.
Second, this relation is used to define the transition relation $\colarrow{\ \ \cdot\ \ }{\cdot}$ which describes the behaviour of collectives.
Finally, the transition relation $\sysarrow{\ \ \cdot\ \ }{}$ will be defined to show how \caspa{} systems evolve.

All these transition relations are defined in the \textsc{FuTS} style~\cite{DLLM13}. 
Using this approach, a transition relation is described using a triple of the form $(\col,\ell,\mathcal{N})$.  The first element of this triple is either  
a component, or a collective, or a system.  The second element is a transition label.  The third element is a function associating each 
component, collective, or system with a non-negative number.  
A non-zero value represents the rate of the exponential distribution characterising the time needed for the execution of the action represented by $\ell$.
The zero value is associated with unreachable terms.  
\modif{We use the  \textsc{FuTS} style semantics because it makes explicit an underlying Action Labelled Markov Chain, which can be simulated 
with standard algorithms~\cite{Gil76} but is nevertheless more compact than Plotkin-style semantics, as the functional form allows different possible
outcomes to be treated within a single rule.}
A complete description of \textsc{FuTS} and their use can be found in~\cite{DLLM13}.
%

%

\subsection{Operational semantics of components}
We use the transition relation ${\comparrow{}{\varepsilon}} \subseteq \cscomps \times \Labs \times [\cscomps\rightarrow \mathbb{R}_{\geq 0}]$ to define the behaviour of a single component.  In this relation $[\cscomps\rightarrow \mathbb{R}_{\geq 0}]$ denotes the set of functions from 
$\cscomps$ to $\mathbb{R}_{\geq 0}$ and $\Labs$ is the set of transition labels $\ell$ which are generated by the following grammar:
\[
\begin{array}{rcll}
\ell & ::= & \labbout{\alpha}{\guard}{\seq{v}}{\csstore} & \mbox{Broadcast output}\\[4pt]
      & \mid     & \labbin{\alpha}{\guard}{\seq{v}}{\csstore}  & \mbox{Broadcast input}\\[4pt]
      & \mid     & \labout{\alpha}{\guard}{\seq{v}}{\csstore} & \mbox{Unicast Output}\\[4pt]
      & \mid     & \labin{\alpha}{\guard}{\seq{v}}{\csstore} & \mbox{Unicast Input}\\[4pt]
      & \mid     & \labsync{\alpha}{\guard}{\seq{v}}{\csstore} & \mbox{Unicast Synchronization}\\[4pt]
      & \mid     & \refusal{\alpha}{\guard}{\seq{v}}{\csstore} & \mbox{Broadcast Input Refusal}\\
\end{array}
\]
The first four labels are associated with the four \caspa{} input-output actions and they contain a reference to the action which is performed ($\alpha$ or $\alpha^{\broadcastsymbol}$), the store of the component where the action is executed~($\csstore$), and
the value which is transmitted or received. The transition label $\labsync{\alpha}{\guard}{\seq{v}}{\csstore}$ is the one which is associated with \emph{unicast synchronisation}.  The final label  $\refusal{\alpha}{\guard}{\seq{v}}{\csstore}$ denotes the case where a component
is not able to receive a broadcast output.  This arises at the level of the single component either because the associated message has been lost, 
or because no process is willing to receive that message. We will observe later in this section that the use of $\refusal{\alpha}{\guard}{\seq{v}}{\csstore}$
labels are crucial to handle appropriately \emph{dynamic process operators}, namely \emph{choice} and \emph{guard}.

The transition relation $\comparrow{}{\varepsilon}$, as formally defined in Table~\ref{tab:compsemantics1} and Table~\ref{tab:compsemantics2}, 
is parametrised with respect to an \emph{evaluation context}
$\varepsilon$. This is used to compute the actual rate of process actions and to 
compute the probability to receive messages.

\begin{table}[tbp]
\begin{center}
$
\begin{array}{cc}
\infer[\rulename{Nil}]{
\cscomp{\csnil}{\csstore} \comparrow{\ell}{\varepsilon} \emptyset
}{
\ell\not=\refusal{\alpha}{\guard}{\seq{v}}{\csstore}
}
&
\infer[\rulename{Nil-F1}]{
\cscomp{\csnil}{\csstore} \comparrow{\refusal{\alpha}{\guard}{\seq{v}}{\csstore}}{\varepsilon} [\cscomp{\csnil}{\csstore}\mapsto 1]
}{
}\\[.5cm]
\multicolumn{2}{c}{
\infer[\rulename{B-Out}]{
\cscomp{\csbout{\alpha}{\guard}{\seq{e}}{\csupd}.P}{\csstore}
\comparrow{\labbout{\alpha}{\guard'}{\seq{v}}{\csstore}}{\varepsilon}
\mu_{r}(\gamma,\alpha^{\broadcastsymbol} )\cdot (P,\mathbf{p})
}{
\llbracket
\pi
\rrbracket_{\gamma}=\pi'
&
\llbracket
\seq{e}
\rrbracket_{\gamma}=\seq{v}
&
\mathbf{p}=\csupd(\csstore)
&
\varepsilon=\langle\mu_p,\mu_r,\mu_u\rangle
}
}\\[.5cm]
\multicolumn{2}{c}{
\infer[\rulename{B-Out-F1}]{
\cscomp{\csbout{\alpha}{\guard_1}{\seq{e}}{\csupd}.P}{\csstore}
\comparrow{\refusal{\beta}{\guard_2}{\seq{v}}{\csstore}}{\varepsilon}
[ \cscomp{\csbout{\alpha}{\guard_1}{\seq{e}}{\csupd}.P}{\csstore} \mapsto 1 ]
}{
} 
}\\[.5cm]
\multicolumn{2}{c}{
\infer[\rulename{B-Out-F2}]{
\cscomp{\csbout{\alpha}{\guard}{\seq{e}}{\csupd}.P}{\csstore}
\comparrow{\ell}{\varepsilon}
\emptyset
}{
\llbracket
\pi
\rrbracket_{\gamma}=\pi'
&
\llbracket
\seq{e}
\rrbracket_{\gamma}=\seq{v}
&
\ell \not=\labbout{\alpha}{\guard'}{\seq{v}}{\csstore}
&
\ell \not=\refusal{\beta}{\guard'}{\seq{v}_1}{\csstore}
}
} \\[.5cm]
\multicolumn{2}{c}{
\infer[\rulename{B-In}]{
\cscomp{\csbin{\alpha}{\guard_2}{\seq{x}}{\csupd}.P}{\csstore_2}
\comparrow{\labbin{\alpha}{\guard_1}{\seq{v}}{\csstore_1}}{\varepsilon}
\mu_{p}(\gamma_1,\gamma_2,\alpha^{\broadcastsymbol} )
\cdot
(P[\seq{v}/\seq{x}],\mathbf{p})
}{
\llbracket
\pi_2[\seq{v}/\seq{x}]
\rrbracket_{\csstore_2}=\guard_2'
&
\csstore_1\models \guard_2'
&
\csstore_2\models\guard_1
&
\mathbf{p}=\csupd[\seq{v}/\seq{x}](\csstore_2)
&
\varepsilon=\langle\mu_p,\mu_r,\mu_u\rangle
}
}\\[.5cm]
\multicolumn{2}{c}{
\infer[\rulename{B-In-F1}]{
\cscomp{\csbin{\alpha}{\guard_2}{\seq{x}}{\csupd}.P}{\csstore_2}
\comparrow{\refusal{\alpha}{\guard_1}{\seq{v}}{\csstore_1}}{\varepsilon}
\left[
\cscomp{\csbin{\alpha}{\guard_2}{\seq{x}}{\csupd}.P}{\csstore_2}\mapsto 1-\mu_{p}(\gamma_1,\gamma_2,\alpha^{\broadcastsymbol} )
\right]
}{
\llbracket
\pi_2[\seq{v}/\seq{x}]
\rrbracket_{\csstore_2}=\guard_2'
&
\csstore_1\models \guard_2'
&
\csstore_2\models\guard_1
&
\varepsilon=\langle\mu_p,\mu_r,\mu_u\rangle
}
}\\[.5cm]
\multicolumn{2}{c}{
\infer[\rulename{B-In-F2}]{
\cscomp{\csbin{\alpha}{\guard_2}{\seq{x}}{\csupd}.P}{\csstore_2}
\comparrow{\labbin{\alpha}{\guard_1}{\seq{v}}{\csstore_1}}{\varepsilon}
\emptyset
}{
\llbracket
\pi_2[\seq{v}/\seq{x}]
\rrbracket_{\csstore_2}=\guard_2'
&
(\csstore_1\not\models \guard_2' 
\mbox{ or }
\csstore_2\not\models\guard_1)
} 
} \\[.5cm]
\multicolumn{2}{c}{
\infer[\rulename{B-In-F3}]{
\cscomp{\csbin{\alpha}{\guard_2}{\seq{x}}{\csupd}.P}{\csstore_2}
\comparrow{\ell}{\varepsilon}
\emptyset
}{
\ell\not=\labbin{\alpha}{\guard_1}{\seq{v}}{\csstore_1}
&
\ell\not=\refusal{\alpha}{\guard_1}{\seq{v}}{\csstore_1}
}
} \\[.5cm]
\multicolumn{2}{c}{
\infer[\rulename{B-In-F4}]{
\cscomp{\csbin{\alpha}{\guard_2}{\seq{x}}{\csupd}.P}{\csstore_2}
\comparrow{\refusal{\beta}{\guard_1}{\seq{v}}{\csstore_1}}{\varepsilon}
[\cscomp{\csbin{\alpha}{\guard_2}{\seq{x}}{\csupd}.P}{\csstore_2}
\mapsto 1]
}{
\alpha\not=\beta
}
}
\end{array}
$
\end{center}
\caption{Operational semantics of components (Part 1)}
\label{tab:compsemantics1}
\end{table}

%

The process $\csnil$ denotes the process that cannot perform any action. The transitions which are induced by this process at
the level of components can be derived via rules \rulename{Nil} and \rulename{Nil-F1}. These rules respectively say that 
the inactive process cannot perform any action, and always refuses any broadcast input.
Note that, the fact that a component $\cscomp{\csnil}{\csstore}$ does not perform any transition is derived from
the fact that any label that is not a \emph{broadcast input refusal} leads to function $\emptyset$ (rule \rulename{Nil}). 
Indeed, $\emptyset$ denotes the $0$ constant function. Conversely, \rulename{Nil-F1} states that $\cscomp{\csnil}{\csstore}$
can always perform a transition labelled $\refusal{\alpha}{\guard}{\seq{v}}{\csstore}$ leading to  $[ \cscomp{\csnil}{\csstore} \mapsto 1]$, 
where $[ \comp \mapsto v]$ denotes the function mapping the component $\comp$ to $v\in \mathbb{R}_{\geq 0}$ and all the other components to $0$.

The behaviour of a \emph{broadcast output} $\cscomp{\csbout{\alpha}{\guard_1}{\seq{e}}{\csupd}.P}{\csstore}$ is described by rules 
\rulename{B-Out}, \rulename{B-Out-F1} and \rulename{B-Out-F2}.
Rule \rulename{B-Out} states that a broadcast output $\csbout{\alpha}{\guard}{\seq{e}}{\csupd}$  can affect components that
satisfy $\pi'=\llbracket\pi \rrbracket_{\gamma}$\footnote{We let $\llbracket\cdot \rrbracket_{\gamma}$ denote the evaluation function of 
an expression/predicate with respect to the store $\csstore$.}. The action rate is determined by the evaluation context 
$\varepsilon=\langle\mu_p,\mu_r,\mu_u\rangle$ and, in particular, by the function $\mu_r$. This function, given a store $\gamma$
and the kind of action performed, in this case $\alpha^{\broadcastsymbol}$, returns a value in $\mathbb{R}_{\geq 0}$. 
If this value is greater than $0$, it denotes the execution rate of the action. However, 
the evaluation context can disable the execution of some actions. This happens when $\mu_{r}(\gamma,\alpha^{\broadcastsymbol})=0$.
The possible next local stores after the execution of an action are determined by the update $\csupd$. This takes the store $\csstore$ and
yields a probability distribution $\mathbf{p}=\csupd(\csstore)\in Dist(\Gamma)$. 
In rule \rulename{B-Out}, and in the rest of the paper, the following notations are used:
\begin{itemize}
\item let $P\in\csprocesses$ and $\mathbf{p}\in Dist(\Gamma)$, $\cscomp{P}{\mathbf{p}}$ is a probability distribution in $Dist(\cscomps)$
such that:
\[
\cscomp{P}{\mathbf{p}}(\comp) = \left\{
\begin{array}{ll}
1 & P\equiv Q|\cskill~\wedge~\comp \equiv\cszero \\
\mathbf{p}(\csstore) & \comp\equiv\cscomp{P}{\csstore}~\wedge~P\not\equiv Q|\cskill\\
0 & \mbox{otherwise}
\end{array}
\right.
\]
\item let $\mathbf{c} \in Dist(\cscomps)$ and $r\in\mathbb{R}_{\geq 0}$, $r\cdot \mathbf{c}$ denotes the function 
$\mathcal{C}:\cscomps \rightarrow\mathbb{R}_{\geq 0}$ such that:
$
\mathcal{C}(\comp) = r\cdot \mathbf{c}(\comp)
$
\end{itemize}

Note that, after the execution of an action a component can be destroyed. This happens when the continuation process after the 
action prefixing contains the term $\cskill$. For instance, by applying rule \rulename{B-Out} we have that: 
$\cscomp{\csbout{\alpha}{\guard_1}{v}{\csupd}.(\cskill|Q)}{\csstore}
\comparrow{\labbout{\alpha}{\guard_1}{v}{\csstore}}{\varepsilon}
[\cszero \mapsto r]$. 

Rule \rulename{B-Out-F1} states that a \emph{broadcast output} always refuses any \emph{broadcast input}, while
\rulename{B-Out-F2} states that a  \emph{broadcast output} can be only involved in labels of the form  $\labbout{\alpha}{\guard}{\seq{v}}{\csstore}$
or $\refusal{\beta}{\guard_2}{\seq{v}}{\csstore}$.

Transitions related to a broadcast input are labelled with $\labbin{\alpha}{\guard_1}{\seq{v}}{\csstore_1}$. 
There, $\csstore_1$ is the store of the component executing the output, $\alpha$ is the action performed,
$\guard_1$ is the predicate that identifies the target components, while $\seq{v}$ is the sequence of transmitted values.
Rule \rulename{B-In} states that a component $\cscomp{\csbin{\alpha}{\guard_2}{\seq{x}}{\csupd}.P}{\csstore_2}$ can perform 
a transition with this label when its store $\csstore_2$ satisfies the target predicate, i.e.\ $\csstore_2\models\guard_1$, and the 
component executing the action satisfies the predicate $\guard_2[\seq{v}/\seq{x}]$.
The evaluation context $\varepsilon=\langle\mu_p,\mu_r,\mu_u\rangle$ can influence the possibility to perform this
action. This transition can be performed with probability  $\mu_{p}(\gamma_1,\gamma_2,\alpha^{\broadcastsymbol} )$.

Rule \rulename{B-In-F1} models the fact that even if a component can potentially receive a broadcast message, the message can get lost
according to a given probability regulated by the evaluation context, namely $1-\mu_{p}(\gamma_1,\gamma_2,\alpha^{\broadcastsymbol} )$. 
Rule \rulename{B-In-F2} models the fact that if a component is not in the set of possible receivers ($\csstore_2\not\models\guard_1$) 
or the sender does not satisfy the expected requirements ($\csstore_1\not\models\guard_2'$) then the component 
cannot receive a broadcast message.
Finally, rules \rulename{B-In-F3} and  \rulename{B-In-F4} model the fact that $\cscomp{\csbin{\alpha}{\guard_2}{\seq{x}}{\csupd}.P}{\csstore_2}$ can only perform a broadcast input on action $\alpha$ and that it always refuses input on any other action type $\beta\not=\alpha$, respectively.

\begin{table}[tbp]
\begin{center}
$
\begin{array}{c}
\infer[\rulename{Out}]{
\cscomp{\csout{\alpha}{\guard}{\seq{e}}{\csupd}.P}{\csstore}
\comparrow{\labout{\alpha}{\guard'}{\seq{v}}{\csstore}}{\varepsilon}
\mu_{r}(\gamma,\alpha)
\cdot
(P,\mathbf{p}) 
}{
\llbracket
\pi
\rrbracket_{\gamma}=\pi'
&
\llbracket
\seq{e}
\rrbracket_{\gamma}=\seq{v}
&
\mathbf{p}=\csupd(\csstore)
&
\varepsilon=\langle\mu_p,\mu_r,\mu_u\rangle
}\\[.5cm]
\infer[\rulename{Out-F1}]{
\cscomp{\csout{\alpha}{\guard_1}{\seq{e}}{\csupd}.P}{\csstore_1}
\comparrow{\refusal{\beta}{\guard_2}{\seq{v}}{\csstore_2}}{\varepsilon}
[
\cscomp{\csout{\alpha}{\guard_1}{\seq{e}}{\csupd}.P}{\csstore_1}\mapsto 1
]
}{
}\\[.5cm]
\infer[\rulename{Out-F2}]{
\cscomp{\csout{\alpha}{\guard}{\seq{e}}{\csupd}.P}{\csstore}
\comparrow{\ell}{\varepsilon}
\emptyset
}{
\llbracket
\pi
\rrbracket_{\gamma}=\pi'
&
\llbracket
\seq{e}
\rrbracket_{\gamma}=\seq{v}
&
\ell \not=\labout{\alpha}{\guard'}{\seq{v}}{\csstore}
&
\ell \not=\refusal{\alpha}{\guard'}{\seq{v}}{\csstore}
}\\[.5cm]
\infer[\rulename{In}]{
\cscomp{\csin{\alpha}{\guard_2}{\seq{x}}{\csupd}.P}{\csstore_2}
\comparrow{\labin{\alpha}{\guard_1}{\seq{v}}{\csstore_1}}{\varepsilon}
\mu_{p}(\gamma_1,\gamma_2,\alpha )
\cdot
(P[\seq{v}/\seq{x}],\mathbf{p})
}{
\llbracket
\pi_2[\seq{v}/\seq{x}]
\rrbracket_{\csstore_2}=\guard_2'
&
\csstore_1\models \guard_2'
&
\csstore_2\models\guard_1
&
\mathbf{p}=\csupd[\seq{v}/\seq{x}](\csstore_2)
&
\varepsilon=\langle\mu_p,\mu_r,\mu_u\rangle
}\\[.5cm]
\infer[\rulename{In-F1}]{
\cscomp{\csin{\alpha}{\guard_2}{\seq{x}}{\csupd}.P}{\csstore_2}
\comparrow{\refusal{\beta}{\guard_1}{\seq{v}}{\csstore_1}}{\varepsilon}
[ \cscomp{\csin{\alpha}{\guard_2}{\seq{x}}{\csupd}.P}{\csstore_2} \mapsto 1 ]
}{
} \\[.5cm] 
\infer[\rulename{In-F2}]{
\cscomp{\csin{\alpha}{\guard_2}{\seq{x}}{\csupd}.P}{\csstore_2}
\comparrow{\labin{\alpha}{\guard_1}{\seq{v}}{\csstore_1}}{\varepsilon}
\emptyset
}{
\llbracket
\pi_2[\seq{v}/\seq{x}]
\rrbracket_{\csstore_2}=\guard_2'
&
(\csstore_1\not\models \guard_2' 
\mbox{ or }
\csstore_2\not\models\guard_1)
}\ \ 
\infer[\rulename{In-F3}]{
\cscomp{\csin{\alpha}{\guard_2}{\seq{x}}{\csupd}.P}{\csstore_2}
\comparrow{\ell}{\varepsilon}
\emptyset
}{
\ell \not=\labin{\alpha}{\guard_1}{\seq{v}}{\csstore_1}
&
\ell\not=\refusal{\beta}{\guard_1}{\seq{v}}{\csstore_1}
}\\[.5cm]
\infer[\rulename{Plus}]{
\cscomp{P+Q}{\csstore}\comparrow{\ell}{\varepsilon} \mathcal{C}_1\oplus\mathcal{C}_2
}{
\cscomp{P}{\csstore} \comparrow{\ell}{\varepsilon} \mathcal{C}_1 &
\cscomp{Q}{\csstore} \comparrow{\ell}{\varepsilon} \mathcal{C}_2 & 
\ell\not=\refusal{\alpha}{\guard'}{\seq{v}}{\csstore}
}\\[.5cm]
\infer[\rulename{Plus-F1}]{
\cscomp{P+Q}{\csstore}\comparrow{\refusal{\alpha}{\guard'}{\seq{v}}{\csstore}}{\varepsilon} \mathcal{C}_1+\mathcal{C}_2
}{
\cscomp{P}{\csstore} \comparrow{\refusal{\alpha}{\guard'}{\seq{v}}{\csstore}}{\varepsilon} \mathcal{C}_1 &
\cscomp{Q}{\csstore} \comparrow{\refusal{\alpha}{\guard'}{\seq{v}}{\csstore}}{\varepsilon} \mathcal{C}_2 & 
}\\[.5cm]
\infer[\rulename{Par}]{
\cscomp{P|Q}{\csstore}\comparrow{\ell}{\varepsilon} \mathcal{C}_1|Q\oplus P|\mathcal{C}_2
}{
\cscomp{P}{\csstore} \comparrow{\ell}{\varepsilon} \mathcal{C}_1 &
\cscomp{Q}{\csstore} \comparrow{\ell}{\varepsilon} \mathcal{C}_2 &
\ell \not=\refusal{\alpha}{\guard}{\seq{v}}{\csstore}
}\\[.5cm]
\infer[\rulename{Par-F1}]{
\cscomp{P|Q}{\csstore}\comparrow{\refusal{\alpha}{\guard}{\seq{v}}{\csstore}}{\varepsilon} \mathcal{C}_1|\mathcal{C}_2
}{
\cscomp{P}{\csstore} \comparrow{\refusal{\alpha}{\guard}{\seq{v}}{\csstore}}{\varepsilon} \mathcal{C}_1 &
\cscomp{Q}{\csstore} \comparrow{\refusal{\alpha}{\guard}{\seq{v}}{\csstore}}{\varepsilon} \mathcal{C}_2
}
\quad
\infer[\rulename{Rec}]{
\cscomp{A}{\csstore}\comparrow{\ell}{\varepsilon} \mathcal{C}
}{
A\defi P &
\cscomp{P}{\csstore} \comparrow{\ell}{\varepsilon} \mathcal{C}
}\\[.5cm]
\infer[\rulename{Guard}]{
\cscomp{ \csguard{\guard}{P}}{\csstore}\comparrow{\ell}{\varepsilon} \mathcal{C}
}{
\csstore \models \guard &
\cscomp{P}{\csstore} \comparrow{\ell}{\varepsilon} \mathcal{C} &
\ell \not=\refusal{\alpha}{\guard}{\seq{v}}{\csstore}
} \quad
\infer[\rulename{Guard-F1}]{
\cscomp{ \csguard{\guard}{P}}{\csstore}\comparrow{\refusal{\alpha}{\guard}{\seq{v}}{\csstore}}{\varepsilon} \csguard{\guard}{\mathcal{C}}
}{
\csstore \models \guard &
\cscomp{P}{\csstore} \comparrow{\refusal{\alpha}{\guard}{\seq{v}}{\csstore}}{\varepsilon} \mathcal{C}
}\\[.5cm]
\infer[\rulename{Guard-F2}]{
\cscomp{ \csguard{\guard}{P}}{\csstore}\comparrow{\ell}{\varepsilon} 
\emptyset
}{
\csstore \not\models \guard &
\ell \not= \refusal{\alpha}{\guard}{\seq{v}}{\csstore}
}\ \ 
\infer[\rulename{Guard-F3}]{
\cscomp{ \csguard{\guard}{P}}{\csstore}\comparrow{\refusal{\alpha}{\guard}{\seq{v}}{\csstore}}{\varepsilon} 
[\cscomp{ \csguard{\guard}{P}}{\csstore}\mapsto 1]
}{
\csstore \not\models \guard 
}
\end{array}
$
\end{center}
\caption{Operational semantics of components (Part 2)}
\label{tab:compsemantics2}
\end{table}

The behaviour of \emph{unicast output} and \emph{unicast input} is defined by the first six rules of Table~\ref{tab:compsemantics2}.
These rules are similar to the ones already presented for broadcast output and broadcast input. The only difference is that 
both unicast output (\rulename{Out-F1}) and unicast input (\rulename{In-F1}) always refuse any broadcast input with probability $1$.
The other rules of Table~\ref{tab:compsemantics2} describe the behaviour of other process operators, namely \emph{choice} $P+Q$, 
\emph{parallel composition} $P|Q$, \emph{guard} and \emph{recursion}. 

The term $P+Q$ identifies a process that can behave either as $P$ or as $Q$. The rule \rulename{Plus} states that the  
components that are reachable by $\cscomp{P+Q}{\csstore}$, via a transition that is not a \emph{broadcast input refusal}, are the ones 
that can be reached either by $\cscomp{P}{\csstore}$ or by $\cscomp{Q}{\csstore}$. In this rule we use $\mathcal{C}_1\oplus \mathcal{C}_2$ to denote the function that maps each term $\comp$ to $\mathcal{C}_1(\comp)+ \mathcal{C}_2(\comp)$, for any $\mathcal{C}_1, \mathcal{C}_2\in[\cscomps\rightarrow \mathbb{R}_{\geq 0}]$. 
At the same time, process $P+Q$ \emph{refuses} a broadcast input when both the process $P$ and $Q$ do that. This is modelled by 
\rulename{Plus-F1}, where, for each $\mathcal{C}_1:\cscomps\rightarrow \mathbb{R}_{\geq 0}$ and 
$\mathcal{C}_2:\cscomps\rightarrow \mathbb{R}_{\geq 0}$,  $\mathcal{C}_1+ \mathcal{C}_2$ denotes the function that maps each term of the form $\cscomp{P+Q}{\csstore}$ to $\mathcal{C}_1(\cscomp{P}{\csstore})\cdot \mathcal{C}_2(\cscomp{Q}{\csstore})$,
while any other component is mapped to $0$. 
Note that, differently from rule \rulename{Plus}, when rule \rulename{Plus-F1} is applied operator $+$ is not removed after the transition. 
This models the fact that when a broadcast message is refused the choice is not resolved. 

In $P|Q$ the two composed processes interleave for all the transition labels except for broadcast input refusal (\rulename{Par}). For this label the two processes synchronise (\rulename{Par-F1}). This models the fact that a message is lost when both  processes refuse to receive it.
In the rules the following notations are used:
\begin{itemize}
\item for each component $\comp$ and process $Q$ we let:
\[
\comp|Q=\left\{
\begin{array}{ll}
\cszero & \comp\equiv\cszero \\
\cscomp{P|Q}{\csstore} & \comp\equiv\cscomp{P}{\csstore} \\
\end{array}
\right.
\]
$Q|\comp$ is symmetrically defined. 
\item for each $\mathcal{C}:\cscomps\rightarrow \mathbb{R}_{\geq 0}$ and process $Q$, $\mathcal{C}|Q$ (resp.\ $Q|\mathcal{C}$) denotes the function that maps each term of the form $\comp|Q$  
(resp.\ $Q|\comp$)
to $\mathcal{C}(\comp)$,
while the others are mapped to $0$;
\item for each $\mathcal{C}_1:\cscomps\rightarrow \mathbb{R}_{\geq 0}$ and $\mathcal{C}_2:\cscomps\rightarrow \mathbb{R}_{\geq 0}$, 
$\mathcal{C}_1| \mathcal{C}_2$ denotes the function that maps each term of the form $\cscomp{P|Q}{\csstore}$ to $\mathcal{C}_1(\cscomp{P}{\csstore})\cdot \mathcal{C}_2(\cscomp{Q}{\csstore})$,
while the others are mapped to $0$.
\end{itemize}

Rule \rulename{Rec} is standard. The behaviour of $\cscomp{ \csguard{\guard}{P}}{\csstore}$ is regulated by rules \rulename{Guard}, \rulename{Guard-F1}, \rulename{Guard-F2} and \rulename{Guard-F3}. The first two rules state that $\cscomp{ \csguard{\guard}{P}}{\csstore}$ behaves exactly like 
$\cscomp{ P}{\csstore}$ when $\csstore$ satisfies predicate $\guard$. However, in the first case the \emph{guard} is removed when a transition is performed.
In contrast, the \emph{guard} still remains active after the transition when a broadcast input is refused. This is similar to what we consider for the
rule \rulename{Plus-F1} and models the fact that broadcast input refusals do not remove \emph{dynamic operators}.
In rule \rulename{Guard-F1} we 
let $\csguard{\guard}{\mathcal{C}}$ denote the function that maps each term of the form $\cscomp{\csguard{\guard}{P}}{\csstore}$ 
to $\mathcal{C}(\cscomp{P}{\csstore}))$ and any other term to $0$, for each $\mathcal{C}:\cscomps\rightarrow \mathbb{R}_{\geq 0}$.
Rules \rulename{Guard-F2} and \rulename{Guard-F3} state that no component can be reached from $\cscomp{ \csguard{\guard}{P}}{\csstore}$ and
all the broadcast messages are refused when $\csstore$ does not satisfy predicate $\guard$.

%
%

\subsection{Operational semantics of collective}

The operational semantics of a \emph{collective} is defined via the transition relation
$\colarrow{}{\varepsilon}\subseteq \csnets \times \Labs\times [\csnets\rightarrow \mathbb{R}_{\geq 0}]$.  This relation is formally defined
in Table~\ref{tab:collsemantics}.  We use a straightforward adaptation of the notations introduced in the previous section. 

\begin{table}[tbp]
\begin{center}
$
\begin{array}{c}
\infer[\rulename{Zero}]{
\cszero\colarrow{\ell}{\varepsilon} \emptyset
}{}
\quad
\infer[\rulename{Comp-B-In}]{
\cscomp{P}{\csstore} \colarrow{\labbin{\alpha}{\guard}{\seq{v}}{\csstore}}{\varepsilon} \frac{\mathcal{N}_1\oplus\mathcal{N}_2}{\oplus\mathcal{N}_1+\oplus\mathcal{N}_2}
}{
\cscomp{P}{\csstore} \comparrow{\labbin{\alpha}{\guard}{\seq{v}}{\csstore}}{\varepsilon} \mathcal{N}_1
&
\cscomp{P}{\csstore} \comparrow{\refusal{\alpha}{\guard}{\seq{v}}{\csstore}}{\varepsilon} \mathcal{N}_2
} \\[.5cm]
\infer[\rulename{Comp}]{
\cscomp{P}{\csstore} \colarrow{\ell}{\varepsilon} \mathcal{N}
}{
\cscomp{P}{\csstore} \comparrow{\ell}{\varepsilon} \mathcal{N}
&
\ell\not=\refusal{\alpha}{\guard}{\seq{v}}{\csstore}
} \quad
\infer[\rulename{B-In-Sync}]{
\col_1\cscpar \col_2 
\colarrow{\labbin{\alpha}{\guard}{\seq{v}}{\csstore}}{\varepsilon}
\mathcal{N}_1\parallel \mathcal{N}_2
}{
\col_1
\colarrow{\labbin{\alpha}{\guard}{\seq{v}}{\csstore}}{\varepsilon}
\mathcal{N}_1
&
\col_2 
\colarrow{\labbin{\alpha}{\guard}{\seq{v}}{\csstore}}{\varepsilon}
\mathcal{N}_2
} \\[.5cm]
\infer[\rulename{B-Sync}]{
\col_1\cscpar \col_2 
\colarrow{\labbout{\alpha}{\guard}{\seq{v}}{\csstore}}{\varepsilon}
(\mathcal{N}_1^{o}\parallel \mathcal{N}_2^{i}) \oplus
(\mathcal{N}_1^{i}\parallel \mathcal{N}_2^{o})
}{
\col_1
\colarrow{\labbout{\alpha}{\guard}{\seq{v}}{\csstore}}{\varepsilon}
\mathcal{N}_1^{o}
&
\col_1
\colarrow{\labbin{\alpha}{\guard}{\seq{v}}{\csstore}}{\varepsilon}
\mathcal{N}_1^{i}
&
\col_2 
\colarrow{\labbout{\alpha}{\guard}{\seq{v}}{\csstore}}{\varepsilon}
\mathcal{N}_2^{o}
&
\col_2 
\colarrow{\labbin{\alpha}{\guard}{\seq{v}}{\csstore}}{\varepsilon}
\mathcal{N}_2^{i}
}\\[.5cm]
\infer[\rulename{Out-Sync}]{
\col_1\cscpar \col_2 
\colarrow{\labout{\alpha}{\guard}{\seq{v}}{\csstore}}{\varepsilon}
\mathcal{N}_1\parallel \col_2 \oplus \col_1\parallel \mathcal{N}_2
}{
\col_1
\colarrow{\labout{\alpha}{\guard}{\seq{v}}{\csstore}}{\varepsilon}
\mathcal{N}_1
&
\col_2 
\colarrow{\labout{\alpha}{\guard}{\seq{v}}{\csstore}}{\varepsilon}
\mathcal{N}_2
}\quad
\infer[\rulename{In-Sync}]{
\col_1\cscpar \col_2 
\colarrow{\labin{\alpha}{\guard}{\seq{v}}{\csstore}}{\varepsilon}
\mathcal{N}_1\parallel \col_2 \oplus \col_1\parallel \mathcal{N}_2
}{
\col_1
\colarrow{\labin{\alpha}{\guard}{\seq{v}}{\csstore}}{\varepsilon}
\mathcal{N}_1
&
\col_2 
\colarrow{\labin{\alpha}{\guard}{\seq{v}}{\csstore}}{\varepsilon}
\mathcal{N}_2
}\\[.5cm]
\infer[\rulename{Sync}]{
\col_1\cscpar \col_2 
\colarrow{\labsync{\alpha}{\guard}{\seq{v}}{\csstore}}{\varepsilon}
\frac{(\mathcal{N}_1^{s}\parallel \col_2)\cdot\oplus\mathcal{N}_1^{i}}{
\oplus\mathcal{N}_1^{i}+
\oplus\mathcal{N}_2^{i}
}
\oplus
\frac{(\col_1\parallel \mathcal{N}_2^{s})\cdot\oplus\mathcal{N}_2^{i}}{
\oplus\mathcal{N}_1^{i}+
\oplus\mathcal{N}_2^{i}
}
\oplus
\frac{(\mathcal{N}_1^{o}\parallel\mathcal{N}_2^{i})}{
\oplus\mathcal{N}_1^{i}+
\oplus\mathcal{N}_2^{i}
}
\oplus
\frac{(\mathcal{N}_1^{i}\parallel\mathcal{N}_2^{o})}{
\oplus\mathcal{N}_1^{i}+
\oplus\mathcal{N}_2^{i}
}
}{
\begin{array}{ccc}
\col_1
\colarrow{\labsync{\alpha}{\guard}{\seq{v}}{\csstore}}{\varepsilon}
\mathcal{N}_1^{s}
&
\col_1
\colarrow{\labout{\alpha}{\guard}{\seq{v}}{\csstore}}{\varepsilon}
\mathcal{N}_1^{o} &
\col_1
\colarrow{\labin{\alpha}{\guard}{\seq{v}}{\csstore}}{\varepsilon}
\mathcal{N}_1^{i}
\\
\col_2 
\colarrow{\labsync{\alpha}{\guard}{\seq{v}}{\csstore}}{\varepsilon}
\mathcal{N}_2^{s}
&
\col_2 
\colarrow{\labout{\alpha}{\guard}{\seq{v}}{\csstore}}{\varepsilon}
\mathcal{N}_2^{o}
&
\col_2 
\colarrow{\labin{\alpha}{\guard}{\seq{v}}{\csstore}}{\varepsilon}
\mathcal{N}_2^{i}
\end{array}
}
\end{array}
$
\end{center}
\caption{Operational semantics of collective}
\label{tab:collsemantics}
\end{table}

Rules \rulename{Zero}, \rulename{Comp-B-In} and \rulename{Comp} describe the behaviour of the single component at the level 
of collective. Rule \rulename{Zero} is similar to rule \rulename{Nil} of Table~\ref{tab:compsemantics1} and states that inactive component $\cszero$ 
cannot perform any action. 
Rule \rulename{Comp-B-In} states that the result of a \emph{broadcast input} of a component at 
the level of \emph{collective} is obtained by combining (summing) the transition at the level of \emph{components} 
labelled $\labbin{\alpha}{\guard}{\seq{v}}{\csstore}$ with the one labelled  $\refusal{\alpha}{\guard}{\seq{v}}{\csstore}$. 
This value is then renormalised to obtain a probability distribution. There we use $\oplus\mathcal{N}$ to denote
$\sum_{\col\in\csnets} \mathcal{N}(\col)$. The renormalisation guarantees a reasonable computation of \emph{broadcast output}  
synchronisation rates (see comments on rule \rulename{B-Sync} below).
Note that each component can always perform a \emph{broadcast input} at the level of collective. However, we are not
able to observe if the message has been received or not. Moreover, thanks to renormalisation, if $\comp \colarrow{\labbin{\alpha}{\guard}{\seq{v}}{\csstore}}{\varepsilon} \mathcal{N}$ then $\oplus\mathcal{N}=1$, i.e. $\mathcal{N}$ is a probability distribution over $\csnets$.
Rule \rulename{Comp} simply states that for the single component $C\not=\cszero$ all the transition labels that are not a \emph{broadcast input},
the relation $\colarrow{\ell}{\varepsilon}$ coincides with the relation $\comparrow{\ell}{\varepsilon}$.

Rules \rulename{B-In-Sync} and \rulename{B-Sync} describe broadcast synchronisation. The former states that two collectives 
$\col_1$ and $\col_2$ that operate in parallel synchronise while performing a
broadcast input. This models the fact that the input can be potentially received by both of the collectives. 
In this rule we let $\mathcal{N}_1\parallel \mathcal{N}_2$ denote the function associating  the
value $\mathcal{N}_1(\col_1)\cdot\mathcal{N}_2(\col_2)$ with each term of the form $\col_1\parallel \col_2$ and $0$ with all the other terms.
We can observe that if $\col \colarrow{\labbin{\alpha}{\guard}{\seq{v}}{\csstore}}{\varepsilon} \mathcal{N}$ then, as we have already 
observed for rule \rulename{Comp-B-In}, $\oplus\mathcal{N}=1$ and $\mathcal{N}$ is in fact a probability distribution over $\csnets$.

Rule \rulename {B-Sync} models the synchronisation consequent of a \emph{broadcast output}  performed at the level of a
collective. For each $\mathcal{N}_1:\csnets\rightarrow \mathbb{R}_{\geq 0}$ and $\mathcal{N}_2:\csnets\rightarrow \mathbb{R}_{\geq 0}$,
$\mathcal{N}_1\oplus \mathcal{N}_2$ denotes the function that maps each term $\col$ to $\mathcal{N}_1(\col)+ \mathcal{N}_2(\col)$.

At the level of collective a transition labelled $\labbout{\alpha}{\guard}{\seq{v}}{\csstore}$ identifies
the execution of a broadcast output. When a collective of the form $\col_1\parallel\col_2$ is considered, the result of these kinds of
transitions must be computed (in the \textsc{FuTS} style) by considering:
\begin{itemize}
\item the broadcast output emitted from $\col_1$, obtained by the transition
$
\col_1
\colarrow{\labbout{\alpha}{\guard}{\seq{v}}{\csstore}}{\varepsilon}
\mathcal{N}_1^{o}
$

\item the broadcast input received  by $\col_1$, obtained by the transition
$
\col_1
\colarrow{\labbin{\alpha}{\guard}{\seq{v}}{\csstore}}{\varepsilon}
\mathcal{N}_1^{i}
$

\item the broadcast output emitted from $\col_2$, obtained by the transition
$
\col_2 
\colarrow{\labbout{\alpha}{\guard}{\seq{v}}{\csstore}}{\varepsilon}
\mathcal{N}_2^{o}
$

\item the broadcast input received  by $\col_2$, obtained by the transition
$
\col_2 
\colarrow{\labbin{\alpha}{\guard}{\seq{v}}{\csstore}}{\varepsilon}
\mathcal{N}_2^{i}
$

\end{itemize}
Note that the first synchronises with the last to obtain $\mathcal{N}_1^{o}\parallel \mathcal{N}_2^{i}$,
 while the second synchronises with the third to obtain $\mathcal{N}_1^{i}\parallel \mathcal{N}_2^{o}$. 
The result of such synchronisations are summed to model the \emph{race condition} between the broadcast outputs 
performed within $\col_1$ and $\col_2$ respectively. 
We have to remark that above $\mathcal{N}_1^{o}$ (resp.\ $\mathcal{N}_2^{o}$) is $\emptyset$ when $\col_1$ (resp.\ $\col_2$)
is not able to perform any broadcast output. Moreover, the label of a broadcast synchronisation is again a \emph{broadcast output}.
This allows further synchronisations in a derivation. Finally, it is easy to see that the total rate of a broadcast synchronisation 
is equal to the total rate of \emph{broadcast outputs}. This means that the number of receivers does not affect the rate of a broadcast
that is only determined by the number of senders.

Rules \rulename{Out-Sync}, \rulename{In-Sync} and \rulename{Sync} control the unicast synchronisation. Rule \rulename{Out-Sync}
states that a collective of the form $\col_1\parallel \col_2$ performs a \emph{unicast output} 
if this is performed either in $\col_1$ or in $\col_2$. This is rendered in the operational semantics
as an interleaving rule, where for each $\mathcal{N}:\csnets\rightarrow \mathbb{R}_{\geq 0}$, $\mathcal{N}\parallel \col_2$ denotes the 
function associating $\mathcal{N}(\col_1)$ with each collective of the form $\col_1\parallel \col_2$ and $0$ with all
other collectives.
Rule \rulename{In-Sync} is similar to \textbf{Out-Sync}. However, it considers \emph{unicast input}.

Finally, rule \rulename{Sync} regulates the \emph{unicast synchronisations} and generates transitions with labels
of the form $\labsync{\alpha}{\guard}{\seq{v}}{\csstore}$. This is the result of a synchronisation between transitions labelled
$\labin{\alpha}{\guard}{\seq{v}}{\csstore}$, i.e.\ an input,  and $\labout{\alpha}{\guard}{\seq{v}}{\csstore}$, i.e.\ an output.

In rule \rulename{Sync}, $\mathcal{N}_k^{s}$, $\mathcal{N}_k^{o}$ and $\mathcal{N}_k^{i}$ denote the result of synchronisation 
($\labsync{\alpha}{\guard}{\seq{v}}{\csstore}$), unicast output ($\labout{\alpha}{\guard}{\seq{v}}{\csstore}$)
and unicast input ($\labin{\alpha}{\guard}{\seq{v}}{\csstore}$) within $\col_k$ ($k=1,2$), respectively.  
%
%
The result of a transition labelled $\labsync{\alpha}{\guard}{\seq{v}}{\csstore}$ is therefore obtained by combining:
\begin{itemize}
\item the synchronisations in $\col_1$ with $\col_2$: $\mathcal{N}_1^{s}\parallel \col_2$;
\item the synchronisations in $\col_2$ with $\col_1$: $\col_1\parallel \mathcal{N}_2^{s}$;
\item the output performed by  $\col_1$ with the input performed by $\col_2$: $\mathcal{N}_1^{o}\parallel \mathcal{N}_2^{i}$;
\item the input performed by  $\col_1$ with the output performed by $\col_2$: $\mathcal{N}_1^{i}\parallel \mathcal{N}_2^{o}$.
\end{itemize}

To guarantee a correct computation of synchronisation rates, the first two addendi are renormalised by considering
inputs performed in $\col_2$ and $\col_1$ respectively. This, on one hand, guarantees that the total rate of 
synchronisation $\labsync{\alpha}{\guard}{\seq{v}}{\csstore}$ does not exceed the \emph{output capacity}, i.e.\ the total rate of 
$\labout{\alpha}{\guard}{\seq{v}}{\csstore}$ in $\col_1$ and $\col_2$. On the other hand, since synchronisation rates are renormalised during the
derivation, it also ensures that parallel composition is associative~\cite{DLLM13}.

\subsection{Operational semantics of systems} 

The operational semantics of systems is defined via the transition relation
$\sysarrow{}{}\subseteq \cssystems \times \Labs\times [\cssystems\rightarrow \mathbb{R}_{\geq 0}]$ that is formally defined
in Table~\ref{tab:csyssemantics}. Only synchronisations are considered at the level of systems.

The first rule is \rulename{Sys-B}. This rule states that a system of the form $\cssys{\col}{(\gamma_{g},\rho)}$ can perform a \emph{broadcast output}
when the collective $\col$, under the environment evaluation $\varepsilon=\langle \mu_r,\mu_p,\mu_u\rangle=\rho(\gamma_{g},\col)$,
can evolve at the level of collective with the label $\labbout{\alpha}{\guard}{\seq{v}}{\csstore}$ to $\mathcal{N}$. After the transition, the global
store is updated and a new collective can be created according to function $\mu_{u}$.
\noindent
In rule \rulename{Sys-B} the following notations are used. For each collective $\col_2$, $\mathcal{N}:\csnets\rightarrow \mathbb{R}_{\geq 0}$,
$\mathcal{S}:\cssystems{}\rightarrow \mathbb{R}_{\geq 0}$  and $\mathbf{p}\in \mathit{Dist}(\Gamma)$ we let 
 $\cssys{\mathcal{N}}{(\mathbf{p},\rho)}$ denote the function mapping each system $\cssys{\col}{(\csstore,\rho)}$ to 
$\mathcal{N}(\col)\cdot \mathbf{p}(\csstore)$.
%
%
%
The second rule is \rulename{Sys} that is similar to \rulename{Sys-B} and regulates unicast synchronisations.

\begin{table}[tbp]
\begin{center}
$
\begin{array}{c}
\infer[\rulename{Sys-B}]{
\cssys{\col}{(\gamma_{g},\rho)}
\sysarrow{\labbout{\alpha}{\guard}{\seq{v}}{\csstore}}{}  
\cssys{\mathcal{N}\parallel \col'}{(\sigma(\gamma_{g}),\rho)}
}{
\rho(\gamma_{g},\col)=\varepsilon=\langle \mu_r,\mu_p,\mu_u\rangle &
\col \colarrow{\labbout{\alpha}{\guard}{\seq{v}}{\csstore}}{\varepsilon} \mathcal{N} &
\mu_u(\gamma_{g},\alpha^{\broadcastsymbol})=(\csupd,\col')
}\\[.5cm]
\infer[\rulename{Sys}]{
\cssys{\col}{(\gamma_{g},\rho)}
\sysarrow{\labsync{\alpha}{\guard}{\seq{v}}{\csstore}}{}  
\cssys{\mathcal{N}\parallel \col'}{(\sigma(\gamma_{g}),\rho)}
}{
\rho(\gamma_{g},\col)=\varepsilon=\langle \mu_r,\mu_p,\mu_u\rangle &
\col 
\colarrow{\labsync{\alpha}{\guard}{\seq{v}}{\csstore}}{\varepsilon}  
\mathcal{N} &
\mu_u(\gamma_{g},\alpha)=(\sigma,\col')
}
\end{array}
$
\end{center}
\caption{Operational Semantics of Systems.}
\label{tab:csyssemantics}
\end{table}

\section{\caspa{} at work}
\label{sec:caspaatwork}

In this section we will use \caspa{} to model a \emph{bike sharing} system~\cite{DeM09,wiki:bikes}. These systems are a recent, and increasingly popular, 
form of public transport in urban areas. As a resource-sharing system with large numbers of independent users altering their behaviour due to pricing and other incentives, they are a simple instance of a collective adaptive system, and hence a suitable case study to exemplify the \caspa{} language.

The idea in a bike sharing system is that bikes are made available in a number of stations that are placed in various 
areas of a city. Users that plan to use a bike for a short trip can pick up a bike at a suitable origin station and return it to any other 
station close to their planned destination. 
One of the major issues in bike sharing systems is the availability and distribution of resources, both in terms of available bikes at the 
stations and in terms of available empty parking places in the stations, where users will park the bikes after using them. 


In our scenario we assume that the city is partitioned in homogeneous zones and that all the 
\emph{stations} in the same zone  can be equivalently used by any user in that zone. 
Below, we let $\{ z_0,\ldots,z_n\}$ be the $n$ zones in the city, each of which contains $k$ parking stations.

Each parking station is modelled in \caspa{} via a component of the form:
\[
\cscomp{~G | R~}{\{\mathsf{zone}=\ell , \mathsf{bikes} = i , \mathsf{slots} = j \}}
\]
where
\begin{itemize}
\item $\mathsf{zone}$ is the attribute identifying the zone where the parking station is located;
\item $\mathsf{bikes}$ is the attribute used to count the number of available bikes;
\item $\mathsf{slots}$ is the attribute containing the total number of parking slots in the parking station.
\end{itemize}

Processes $G$ and $R$, which model the procedure to \emph{get} and \emph{return} a bike in the parking station, respectively, are defined as follow:
\[
\begin{array}{rcl}
G & \defi  &\csguard{\mathsf{bikes}>0}{~\csout{\mathsf{get}}{\mathsf{zone}=\mathsf{this}.\mathsf{zone}}{\bullet}{\{\mathsf{bikes} \leftarrow \mathsf{bikes}-1\}}.G}\\[.5cm]

R & \defi  &\csguard{\mathsf{slots}>\mathsf{bikes}}{~\csout{\mathsf{ret}}{\mathsf{zone}=\mathsf{this}.\mathsf{zone}}{\bullet}{\{\mathsf{bikes} \leftarrow \mathsf{bikes}+1\}}.R}
\end{array}
\]

Process $G$, when the value of attribute $\mathsf{bikes}$ is greater than $0$, executes the \emph{unicast output} with action type $\mathsf{get}$ 
that potentially involves components satisfying the predicate $\mathsf{zone}=\mathsf{this}.\mathsf{zone}$, i.e.\ the ones that are located in the same zone\footnote{Here we use $\bullet$ to denote the unit value.}.
When the output is executed the value of the attribute $\mathsf{bikes}$ is decreased by one to model the fact that one bike has been retrieved from 
the parking station.

Process $R$ is similar. It executes the \emph{unicast output} with action type $\mathsf{ret}$ 
that potentially involves components satisfying predicate $\mathsf{zone}=\mathsf{this}.\mathsf{zone}$. This action can be executed only 
when there is at least one parking slot available, i.e.\ when the value of attribute $\mathsf{bikes}$ is less than the value of attribute $\mathsf{slots}$.
When the output considered above is executed, the value of attribute $\mathsf{bikes}$ is increased by one to model the fact that one
bike has been returned in the parking station.


Users, who can be either \emph{bikers} or \emph{pedestrians}, are modelled via components of the form:
\[
\cscomp{ Q }{\{ \mathsf{zone}=\ell \}}
\]
where $\mathsf{zone}$ is the attribute indicating where the user is located, while $Q$ models the current state of the user and can be one of the following
processes:
\[
\begin{array}{rcl}
B & \defi & \csbout{\mathsf{move}}{\bot}{\bullet}{\{\mathsf{zone} \leftarrow U(z_0,\ldots, z_n)\}}.B\\
   & + & \csbout{\mathsf{stop}}{\bot}{\bullet}{}.WS
\\[5pt]
WS & \defi & \csin{\mathsf{ret}}{\mathsf{zone}=\mathsf{this}.\mathsf{zone}}{\bullet}{}.P
\\[5pt] 
P & \defi & \csbout{\mathsf{go}}{\bot}{\bullet}{}.WS
\\[5pt] 
WB & \defi & \csin{\mathsf{get}}{\mathsf{zone}=\mathsf{this}.\mathsf{zone}}{\bullet}{}.B
\end{array}
\]
Process $B$ represents a \emph{biker}. When a user is in this state (s)he can either \emph{move} from the current zone to another zone 
or \emph{stop} to return the bike to a parking station.  These activities are modelled with the execution of a broadcast output via action types $\mathsf{move}$
and $\mathsf{stop}$, respectively. Note that in both of these cases, the predicate used to identify the target of the actions is $\bot$, denoting
the value \emph{false}. This means that neither of the two actions actually synchronise with any component (since no component satisfies $\bot$).
This kind of interaction is used in \caspa{} to model \emph{spontaneous actions}, i.e.\ actions that render the execution of an activity and that
do no require synchronisation. 
After the broadcast $\mathsf{move}^{\broadcastsymbol}$ the value of attribute $\mathsf{zone}$ is updated by randomly selecting the
next zone in $\{z_0,\ldots, z_n\}$. With $\{\mathsf{zone} \leftarrow U(z_0,\ldots, z_n)\}$ we denote the update $\csupd$ such that 
$\csupd(\csstore)$ is the probability distribution giving probability $\frac{1}{n}$ to each store $\csstore[\mathsf{zone} \leftarrow z_i]$. 
This update models a random movement of the user among the city zones.

When process $B$ executes broadcast $\mathsf{stop}^{\broadcastsymbol}$, it evolves to process $WS$. This process models a user 
who is waiting for a parking slot. This process executes an input over $\mathsf{ret}$. This models the fact that the user has found a parking 
station with an available parking slot in their zone. After the execution of this input process $P$ is executed. 
The latter component definition models a \emph{pedestrian user}. The user remains in this state until the \emph{spontaneous action} $\mathsf{go}^{\broadcastsymbol}$
is performed. After that it evolves to process $WB$ which models a user waiting for a bike. The behaviour of $WB$ is similar to that of $WS$ described 
above. 


\begin{figure}[tbp]
\begin{center}
\includegraphics[scale=0.25]{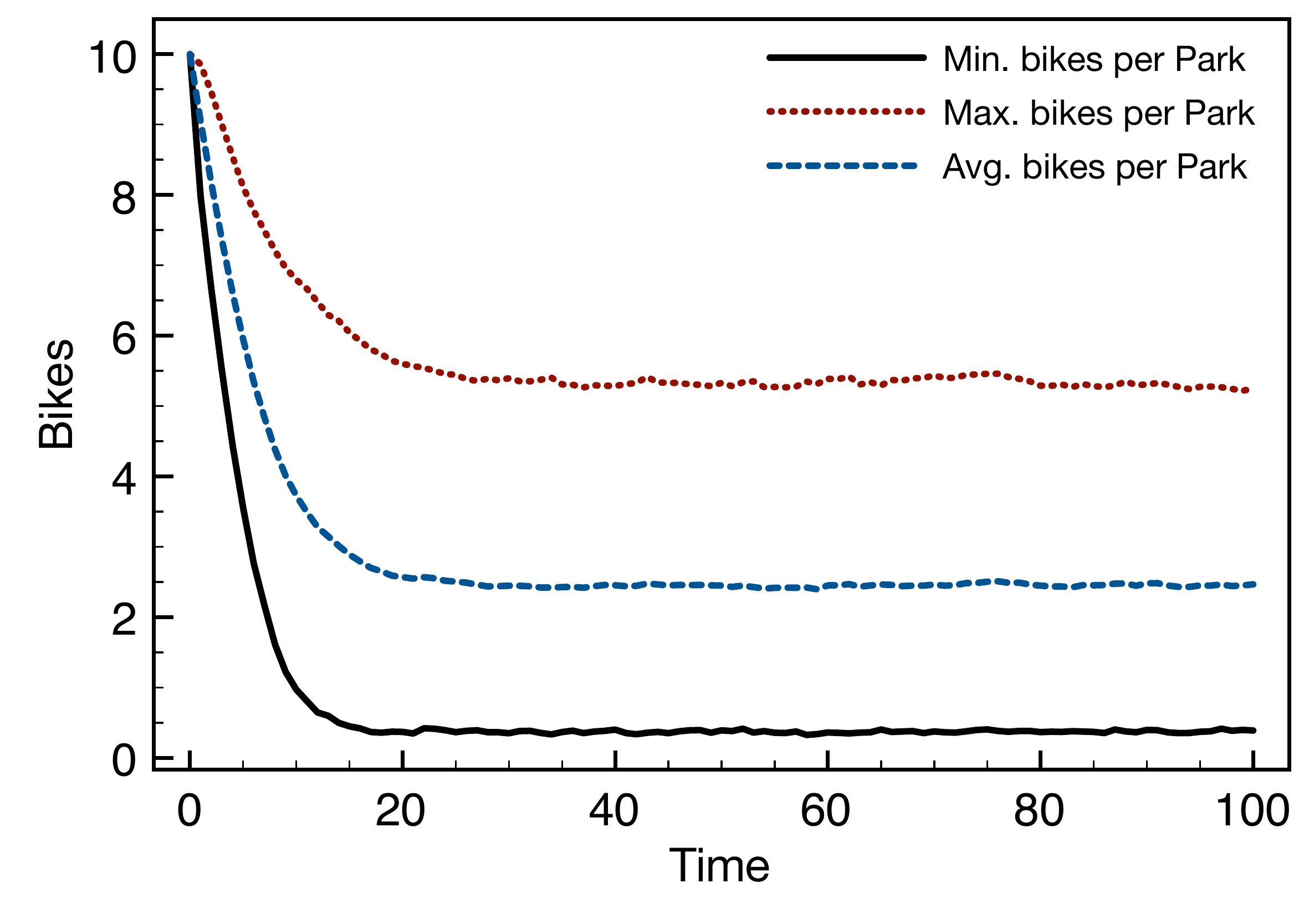} 
\end{center}
\caption{Simulation of bike scenario.}
\label{fig:simulation}
\end{figure}

\modif{Using a custom-built prototype simulator, we are able to simulate this modelled scenario.  The output on one simulation run is presented in Figure~\ref{fig:simulation}. 
In the graph we show the minimum, average and maximum number of
bikes in one zone of the city. We consider a scenario with four zones each containing four parking stations. The total number of users is $150$.}

\section{Conclusions}
\label{sec:conclusions}

We have presented \caspa{}, a new stochastic process algebra for the representation of systems developed
in the CAS paradigm.   The language offers a rich set of communication primitives, and the use of attributes, captured in a store associated with each component, allows attribute-based communication.  For most CAS systems we anticipate that one of the attributes will be the location of the agent and thus it is straightforward to capture systems in which, for example, there is a limited scope of communication, or restriction to only interact with components that are co-located.  As demonstrated in the case study presented in Section~4, attributes can also be used to capture the "state" of a component, such as the available number of bikes/slots at a bike station.

\modif{\caspa{} reflects the experience that we have gained through earlier languages such as SCEL~\cite{NicolaLPT14}, its Markovian variants~\cite{LLMS14} and PALOMA~\cite{palomaqest14}.  Compared with SCEL, the representation of knowledge here is more abstract, and not designed for detailed reasoning during the evolution of the model.  This reflects the different objectives of the languages. Whilst SCEL is designed to support the programming of autonomic computing systems, the primary focus of \caspa{} is quantitative analysis.  In stochastic process algebras such as PEPA, MTIPP and EMPA, data is typically abstracted away, and the influence of data on behaviour is captured only stochastically.  When the data is important to differentiate behaviour it must be implicitly encoded in the state of components.  In the context of CAS we wish to support attribute-based communication to reflect the flexible and dynamic interactions that occur in such systems.  Thus it is not possible to entirely abstract from data.  On the other hand, the level of abstraction means that choices within the system will be captured stochastically rather than through the rich policies for reasoning offered by SCEL\@.  We believe that this offers a reasonable compromise between expressiveness and tractability.  Another key feature of \caspa{} is the inclusion of an explicit environment in which components interact.  In PALOMA there was a rudimentary form of environment, termed the \emph{perception function} but this proved cumbersome to use, and it could not itself be influenced by the behaviour of the components.  In \caspa{}, in contrast, the environment not only modulates the rates and probabilities related to interactions between components, it can also itself evolve at runtime, due to feedback from the collective.}

The focus of this paper has been the discrete semantics in the structured operational style of FUTS \cite{DLLM13},
but in future work we plan to develop differential semantics in the style of \cite{tribastone2012scalable}.   This latter approach
will be essential in order to support quantitative analysis of CAS systems of realistic scale, but it may not
be possible to encompass the full rich set of language features of \caspa{} with such efficient analysis. Further work
is needed to investigate this issue, and which language features can be supported for the various forms of quantitative analysis available.
Additional work involves the development of an appropriate high-level language for designers of CAS which will be mapped to the process algebra, 
and hence will enable qualitative and quantitive analysis of CAS during system development by enabling a design workflow and analysis pathway.
\modif{The intention of this high-level language is not to add to the expressiveness of \caspa{}, which we believe to be well-suited to capturing
the behaviour of CAS, but rather to ease the task of modelling for users who are unfamiliar with process algebra and similar formal notations.}

\section*{Acknowledgements}
This work is partially supported by the EU project QUANTICOL, 600708.  This research has also been partially funded by  the German Research
   Council (DFG) as part of the Cluster of Excellence
   on Multimodal Computing and Interaction at Saarland University.

\bibliographystyle{eptcs}
\bibliography{carma}

\end{document}